\documentclass{elsart3}
\usepackage{epsfig,psfrag,amsmath,amssymb,float}

\begin{document}
\newcommand{\bc}{\begin{center}}
\newcommand{\ec}{\end{center}}
\newcommand{\be}{\begin{equation}}
\newcommand{\ee}{\end{equation}}
\newcommand{\beqn}{\begin{eqnarray}}
\newcommand{\eeqn}{\end{eqnarray}}

\begin{frontmatter}


\title{Magnetic ordering in doped coupled frustrated  quantum  spin-$\frac{1}{2}$ chains with 4-spin exchange.}
\author{Nicolas Laflorencie and Didier Poilblanc}


\address{Laboratoire de Physique Th\'eorique; IRSAMC; 
Universit\'e Paul Sabatier;\\ 31062 Toulouse; France}

\begin{abstract}
The role of various magnetic inter-chain couplings has been investigated recently  by numerical methods in doped frustrated quantum spin chains. A non-magnetic dopant introduced in a gapped spin chain releases a free spin-1/2 soliton. The formation of a local magnetic moment has been analyzed in term of soliton confinement. A four-spin coupling which might originate from cyclic exchange is shown to produce such a confinement. Dopants on different chains experience an effective space-extended non-frustrating pairwise spin interaction. This effective interaction between impurity-spins is long-ranged and therefore is expected to play a crucial role in the mecanism of antiferromagnetic (AF) long-range ordering (LRO) observed in spin-Peierls (SP) compounds such as CuGeO$_3$ doped with non-magnetic impurities.

\end{abstract}

\end{frontmatter}

\section{Introduction}
\label{intro}
\begin{figure}
\centering
\psfrag{J1}{{\small{$J_1$}}} \psfrag{alpha J1}{{\small{$\alpha J_1$}}}
\psfrag{Jp}{{\small{$J_{\bot}$}}} \psfrag{J4}{{\small{$J_4$}}}
\includegraphics[height=3cm]{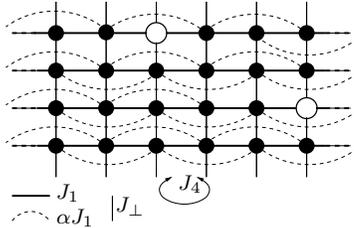}
\caption{Schematic picture
of the coupled chains model with nearest neighbor (NN), next-nearest
neighbor(NNN), inter-chain and $4$-spin couplings $J_1$, $J_2=\alpha
J_1$, $J_{\bot}$, and $J_4$. Full (resp. open) circles stand for
spin-$\frac{1}{2}$ sites (resp. non-magnetic dopants).}
\label{fig:Lattice}
\end{figure}
Doping a spin liquid system with non-magnetic impurities leads to very surprising new features. For example in the doped quasi one-dimensional compound Cu$_{1-x}$M$_x$GeO$_3$ (M$=$Zn or Mg), the discovery of coexistence between dimerization and AF LRO at small impurity concentration $x$ has motivated extented experimental~\cite{Zn_CuGeO} and theoretical~\cite{Sorensen98,Nakamura99,Martins96,Normand2002,Hansen99,Dobry98} investigations. The impurity-induced AF LRO has been observed in other doped spin liquid materials such as the 2-legs ladder Sr(Cu$_{1-x}$Zn$_x$)$_2$O$_3$~\cite{azuma97}, the Haldane compound Pb(Ni$_{1-x}$Mg$_x$)$_2$V$_2$O$_8$~\cite{uchiyama99} or the coupled spin dimer system TlCu$_{1-x}$Mg$_x$Cl$_3$~\cite{oosawa03}.

Replacing a spin-$\frac{1}{2}$ in a {\it spontaneously} dimerised (isolated) spin chain by a non magnetic dopant (described as an inert site) liberates a free spin $\frac{1}{2}$, named a soliton, which does not bind to the dopant~\cite{Sorensen98}. On the other hand, a {\it static} bond dimerisation produces an attractive potential between the soliton and the dopant~\cite{Sorensen98,Nakamura99} and consequently leads, under doping, to the formation of local magnetic moments~\cite{Sorensen98,Normand2002} as well as a rapid suppression of the spin gap~\cite{Martins96}. However, a coupling to a purely one-dimensional (1D) adiabatic lattice~\cite{Hansen99} does not produce confinement in contrast to more realistic models including an elastic inter-chain coupling (to mimic 2D or 3D lattices)~\cite{Hansen99,Dobry98}. 

Frustration and inter-chains effects are necessary to understand the impurity-induced AF ordering in the doped spin-Peierls (SP) material  Cu$_{1-x}$M$_x$GeO$_3$. In section 2 we report numerical studies of models for doped coupled spin chains~\cite{ourprl03} and concentrate on the local moment formation induced by the doping. Dopants on different chains experience an effective space-extended non-frustrating pairwise spin interaction which is long-ranged and therefore is expected to play a crucial role in the mecanism of AF LRO. In section 3, we report exact diagonalisation (ED) results for the effective magnetic coupling which appears between released spins. This long-distance interaction between impurity-spins is finally included in an effective 2D model with a small concentration $x$ of spins-$\frac{1}{2}$ put at random on a square lattice. Concluding remarks are given in Section 4 where we also mention some preliminary results obtained by a Quantum Monte Carlo (QMC) study of an effective diluted model.


\section{Impurity induced local moment formation in doped coupled frustrated spin chains}

Let us first consider a model of coupled frustrated
spin-$\frac{1}{2}$ antiferromagnetic chains (see Fig.\ref{fig:Lattice}). Following
Schulz~\cite{Schulz96}, a mean-field (MF) treatment of the inter-chain
couplings has been performed~\cite{ourprl03} and the resulting Hamiltonian is given by
\begin{eqnarray}
\label{hamilQ1D.MF} H_{\rm eff}(\alpha,J_{\bot},J_4)
=J\sum_{i,a}[(1+\delta J_{i,a})\vec S_{i,a}\cdot \vec S_{i+1,a}
\nonumber \\ + \alpha\vec S_{i,a}\cdot \vec
S_{i+2,a}+h_{i,a}S_{i,a}^z + {\rm constant}\, ,
\end{eqnarray}
where
\begin{equation}
\label{Hi}
h_{i,a}=J_{\bot}(\langle S_{i,a+1}^z \rangle+\langle S_{i,a-1}^z \rangle)
\end{equation}
accounts for first-order effects in the inter-chain magnetic coupling $J_{\perp}$, and
\begin{equation}
\label{delti} \delta J_{i,a}=J_4\lbrace\langle \vec
S_{i,a+1}\cdot\vec S_{i+1,a+1} \rangle+\langle \vec
S_{i,a-1}\cdot\vec S_{i+1,a-1} \rangle\rbrace
\end{equation}
takes a generic form because it might have multiple origins; although a four-spin cyclic exchange~\cite{cyclic,Laeuchli2002} provides the most straightforward derivation of it~\cite{ourprl03}, $J_4$ can also mimic higher order effects in $J_{\perp}$~\cite{Byrnes99} or the coupling to a 2D (or 3D) lattice~\cite{Hansen99}.
\begin{figure}
\begin{center}
\psfrag{A1}{{\small{$J_4=0.01$}}}
\psfrag{A2}{{\small{$J_4=0.05$}}} \psfrag{A3}{{\small{$J_4=0.1$}}}
\psfrag{A4}{{\small{$J_4=0.2$}}} \psfrag{A5}{$J_{\bot}$}
\psfrag{AF}{AF} \psfrag{SP}{SP} \psfrag{ac}{\small{$\alpha_c$}}
\psfrag{alp}{$\alpha$} \psfrag{L12}{\small{$L=12$}}
\psfrag{L16}{\small{$L=16$}} \psfrag{F S E}{{\small {F S E}}}
\epsfig{file=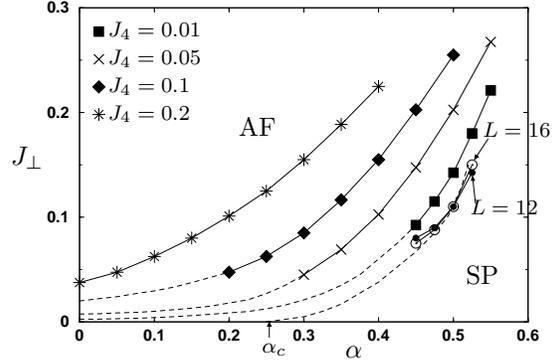,width=7cm} \caption{SP-AF phase diagram in
the ($\alpha,J_{\bot}$) plane from ED of chains of length up to
$16$ sites. Symbols correspond to different values of $J_4\ge 0$
as indicated on plot. Typically, FSE are smaller than the size of
the symbols. The computed transition lines are extended by {\it
tentative} transition lines (dashed lines) in the region where FSE
become large. At $J_4=0$ we have plotted a few points in the
vicinity of the MG point~\cite{MG} for $L=12$ and $L=16$ (reprinted from Ref.\cite{ourprl03}).} 
\label{fig:PhDg2}
\end{center}
\end{figure}
$i$ is a lattice index along the chain of size $L$ and $a$
labels the $M$ chains ($L$ and $M$ chosen to be even). Periodic
boundary conditions will be assumed in {\it both directions}. The energy scale is set by the coupling along the chains $J=1$ and $\alpha$ is the relative magnitude of the NNN frustrating coupling.

In the pure case (i.e. without impurity), all the chains are equivalent and the problem is therefore reduced to a single chain problem in a staggered magnetic field $h_i=-2J_{\perp}<S_{i}^{z}>$ and with its NN exchange modulated by $\delta J_i =J_4\langle \vec S_{i}\cdot\vec S_{i+1} \rangle$ if $J_4<0$ or $\delta J_i =J_4\langle \vec S_{i+1}\cdot\vec S_{i+2} \rangle$ if $J_4>0$. Using Lanczos ED up to the convergence of the MF procedure~\cite{springer03}, we can identify two different phases in the $(\alpha,J_{\perp})$ plane. A dimerised SP phase and an AF ordered phase separated by a transition line $J_{\perp}=J_{\perp}^{c}(\alpha)$ (see Fig.\ref{fig:PhDg2}). ED have been performed on small systems ($L\le 16$) for different values of $J_4$. Fortunately, the finite size effects (FSE) are small in the gapped regime. The modulation created by $J_4$ stabilizes the SP phase, as we can observe on Fig.\ref{fig:PhDg2}

\begin{figure}
\begin{center}
\psfrag{12}{$L=12$} \psfrag{L16}{$L=16$}
\psfrag{S}{{\small{$\langle S_{i,1}^z \rangle$}}}
\psfrag{z}{{\small{$i$}}} \psfrag{J4}{$J_{4}$} \psfrag{xi}{$\xi_{\parallel}$}
\psfrag{lo}{{\small{$2\xi_{\parallel}$}}} \epsfig{file=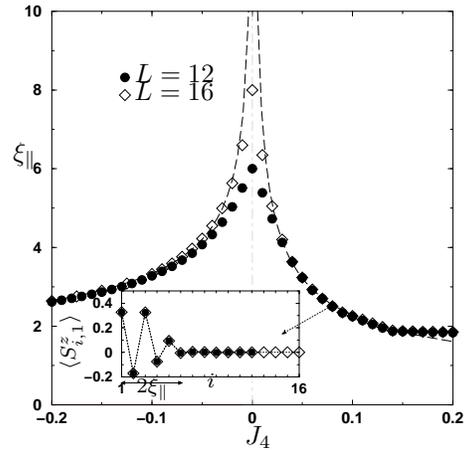,width=6cm}
\caption{ED data of the soliton average position  vs $J_4$
calculated for $\alpha=0.5$ and $J_{\bot}=0.1$. Different symbols
are used for $L\times M$ = $12\times 6$ and $16\times 8$ clusters.
The long-dashed line is a power-law fit (see text). Inset shows
the magnetization profile in the doped ($a=1$) chain at $J_4=0.08$, ie
$\xi_{\parallel} \simeq 2.5$ (reprinted from Ref.\cite{ourprl03}).}
\label{fig:xi}
\end{center}
\end{figure}

Let us now turn to the doped case. A non-magnetic dopant is
described here as an inert site decoupled from its neighbors.
Under doping the system becomes
non-homogeneous so that we define a local mean staggered
magnetization,
\begin{equation}
\label{MeStMg} {\mathcal{M}}^{\rm stag}_{i,a}=\frac{1}{4}
(-1)^{i+a}(2\langle S^{z}_{i,a} \rangle - \langle S^{z}_{i+1,a}
\rangle - \langle S^{z}_{i-1,a} \rangle).
\end{equation}
Following the method used in Ref.~\cite{Dobry98}, the MF equations
are solved self-consistently on finite $L\times M$ clusters and
lead to a non-uniform solution. At each step of the MF iteration
procedure, we use Lanczos ED techniques to treat {\it exactly}
(although independently) the $M$ {\it non-equivalent} finite
chains  and compute
$\langle S^{z}_{i,a}\rangle$ for the next iteration step
until the convergence is eventually achieved.
\begin{figure}
\centering
\psfrag{M1}{${\mathcal{M}}_{i,a}^{\rm stag}$} \psfrag{x}{$i$}
\psfrag{a}{{\small{$J_4=0$}}} \psfrag{b}{{\small{$J_4=0.01$}}}
\psfrag{c}{{\small{$J_4=0.08$}}}
\includegraphics[height=5cm]{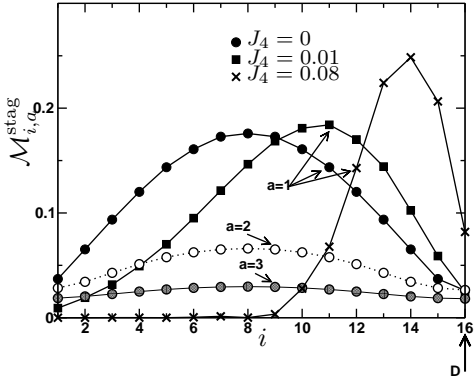} 
\caption{Local
magnetization ${\mathcal{M}}^{\rm stag}_{i,a}$ for $L\times
M$=$16\times 8$ coupled chains with one dopant D (shown by arrow)
located at $a=1$, $i=16$ in the dimerised phase ($\alpha=0.5$,
$J_{\bot}=0.1$). Circles correspond to $J_4=0$ (shown up to the
third neighbor chain of the doped one) and squares (crosses) to
$J_4=0.01$ ($J_4=0.08$). The coupling $J_2$ across the dopant has been
set to $0$ for convenience (reprinted from Ref.\cite{ourprl03}).} 
\label{fig:ConfinJ2p}
\end{figure}
We first consider the case of a single dopant. Whereas in the case $J_4 =0$  the soliton remains de-confined as can be seen from
Fig.\ref{fig:ConfinJ2p}, a very small $J_4 \neq 0$ is sufficient to produce a confining string which binds the soliton to the dopant. Note that the inter-chain coupling
induces a ''polarization cloud'' with strong antiferromagnetic
correlations in the neighbor chains of the doped one; we can therefore define a typical length scale in the transverse direction $\xi_{\perp}$ which is $\simeq 1$ in the case $J_{\perp}=0.1$, as we will study in the last part of next section. A confinement length in the chain direction $\xi_{\parallel}$ can also be extracted. Defined by 
\begin{equation}
\xi_{\parallel}=\frac{\sum_{i}i|S_{i}^{z}|}{\sum_{i}|S_{i}^{z}|},
\end{equation}
we have calculated it for a $16 \times 8$ system with $\alpha=0.5$ and $J_{\perp}=0.1$, and we show its variation as a function of $J_4$ in Fig.\ref{fig:xi}. FSE decrease
for increasing $J_4$.  Note that $\xi_{\parallel}(J_4)\neq\xi_{\parallel}(-J_4)$ and a
power law~\cite{Nakamura99} with different exponents $\eta$ is
expected when $J_4\rightarrow 0$. A fit gives $\eta\sim 0.33$ if
$J_4<0$ and $\eta\sim 0.50$ for $ J_4>0$ (Fig.\ref{fig:xi}). This
asymmetry can be understood from opposite renormalisations of
$J_1$ for different signs of $J_4$. Indeed, if $J_4<0$ then
$\delta J_{i,a}>0$ and the nearest neighbor MF exchange becomes
larger than the bare one. Opposite effects are induced by $J_4>0$.

\section{Effective interaction between impurity-spins}
\begin{figure}
\centering

\includegraphics[height=4cm]{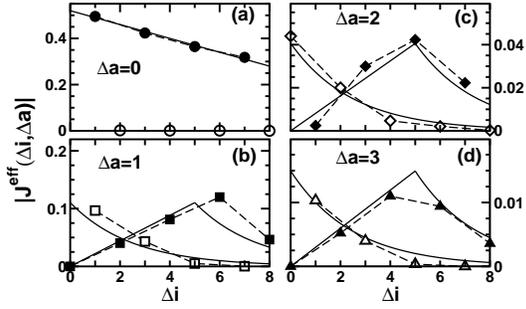} \caption{Magnitude of the
effective magnetic coupling between two impurities located either
on the same chain (a)  or on different ones (b-c-d)  vs the dopant separation $\Delta i$ in
a system of size $L \times M=16 \times 8$ with $\alpha=0.5$,
$J_{\bot}=0.1$, and $J_4=0.08$. Closed (resp. open) symbols
correspond to AF (F) interactions. Full lines are fits (see text)}
\label{fig:Jeff}
\end{figure}

We now turn to the investigation of the effective interaction between dopants. Each impurity releases an effective spin $\frac{1}{2}$, localized at a distance $\sim \xi_{\parallel}$ from it due to the confining potentiel set by $J_4$. 
 We define an effective pairwise interaction $J^{\rm eff}$ as the
energy difference of the $S=1$ and the $S=0$ GS. When $J^{\rm
eff}=E(S=1)-E(S=0)$ is positive (negative) the spin interaction is
AF (ferromagnetic). Let us first consider the case of two
dopants in the same chain. (i) When the two vacancies are on the
same sub-lattice the moments experience a very small ferromagnetic
$J^{\rm eff}<0$ as seen in Fig.~\ref{fig:Jeff} with $\Delta a=0$ so that the
two effective spins $\frac{1}{2}$ are almost free. (ii) When the
two vacancies sit on different sub-lattices, $\Delta i$ is odd and
the effective coupling is AF with a magnitude close to the
singlet-triplet gap. Fig.~\ref{fig:Jeff} with $\Delta a=0$ shows that the decay
of $J^{\rm eff}$ with distance is in fact very slow for such a
configuration. The behavior of the pairwise interaction of two dopants located on
{\it different} chains ($\Delta a=1,2,3$) is shown in
Fig.~\ref{fig:Jeff} for  $\Delta a=1,2,3$~for $J_4>0$. When dopants are on
opposite sub-lattices the effective interaction is
antiferromagnetic. At small dopant separation $J^{\rm eff}(\Delta
i)$ increases with the dopant separation as the overlap between
the two AF clouds increases until $\Delta i \sim 2\xi_{\parallel}$. For larger
separation, $J^{\rm eff}(\Delta i)$ decays rapidly. If dopants are on the same sub-lattice,
solitons are located on the same side of the dopants~\cite{note3}
and the effective exchange $J^{\rm eff}(\Delta i)$ is
ferromagnetic and decays rapidly to become negligible when $\Delta i >
2\xi$. The key feature here is the fact that the
effective pairwise interaction is {\it not} frustrating (because
of its sign alternation with distance) although the frustration is
present in the microscopic underlying model. AF ordering is then
expected (at $T=0$) as seen for a related system of coupled
Spin-Peierls chains~\cite{Dobry98}. 

Our next step is to fit the numerical data in order to derive an analytic expression for $J_{\rm{eff}}$ and a long ranged non frustrationg effective model for diluted spins-$\frac{1}{2}$ on a $L_x\times L_y$ square lattice,
\begin{equation}
\label{eff}
\mathcal{H}^{\rm{eff}}=\sum_{\bf {r}_1 ,\bf {r}_2} \epsilon _{\bf {r}_1}\epsilon _{\bf {r}_2}J_{\rm{eff}}(\bf {r}_1 -\bf {r}_2)\vec
S_{\bf {r}_1}\cdot\vec S_{\bf {r}_2},
\end{equation}
with $\epsilon _{\bf {r}}=1~(0)$ with probability $x~(1-x)$, $x$ being the dopant concentration. Using only five parameters, two energy scales and three length scales, we can fit ED data with very simple mathematical expressions. When $\Delta a=0$ (same chain), $J^{\rm {eff}}$ approximately fulfills $J^{\rm eff}(\Delta i,0)=J_0(1-\Delta i/\xi_{\parallel}^{0})$ for $\Delta i$ even and $\Delta i < \xi_{\parallel}^{0}$, and $J^{\rm eff}(\Delta i,0)=0$ otherwise. For dopants located on different chains and on the same sub-lattice ($\Delta i+\Delta a$ even) one has,
\be
J^{\rm eff}(\Delta i,\Delta a)=-J^{'}_{0}\exp(-\frac{\Delta i}{\xi_{\parallel}})\exp(-\frac{\Delta a}{\xi_{\perp}}),
\ee
while if the dopants are on opposite sub-lattices, one gets
\be
J^{\rm eff}(\Delta i,\Delta a)=J^{'}_{0}\frac{\Delta i}{2\xi_{\parallel}}\exp(-\frac{\Delta a}{\xi_{\perp}})
\ee
for $\Delta i\le 2\xi_{\parallel}$ and 
\be
J^{\rm eff}(\Delta i,\Delta a)=-J^{'}_{0}\exp(-\frac{\Delta i-2\xi_{\parallel}}{\xi_{\parallel}})\exp(-\frac{\Delta a}{\xi_{\perp}}),
\ee
for $\Delta i > 2\xi_{\parallel}$. The fitting parameters are $J_{0}=0.52$, $J_{0}^{'}=0.3$, $\xi_{\parallel}^{0}=17.33$, $\xi_{\parallel}=2.5$ and $\xi_{\perp}=1$ in the case considered here : $\alpha=0.5, J_{\perp}=0.1~{\rm{and}}~J_4 =0.08$ (see Fig.\ref{fig:Jeff}).

\section{Conclusion}
We can conclude this study by mentionning some preliminary results obtained by the way of QMC simulations~\cite{sse} performed on the effective diluted model Eq.(\ref{eff}) with a great number of spins $N_S \le 256$ . Even at very small concentrations $x$, a N\'eel type AF LRO at $T=0$ is observed as a result of the simulations; details about this study will be reported elsewhere~\cite{new}.\\
We gratefully acknowledge Anders W. Sandvik for the interest he took in this work.


\begin{thebibliography}{00}
\bibitem{note3} Two equivalent symmetric solutions are obtained with
opposite signs of the bulk dimerisation.
\end{thebibliography}
\end{document}